# An Analysis of Capital Market through the Lens of Integral Transforms: Exploring Efficient Markets and Information Asymmetry.

**Kiran Sharma**
*Research scholar*
*Department of Commerce, Sikkim University*
*Email: timshinakiran1509@gmail.com*

*Prof Abhijit Dutta*
*Professor, Department of Commerce*
*Sikkim University*
*E-mail: adutta@cus.ac.in*

*Dr Rupak Mukherjee*
*Assistant Professor, Department of Physics*
*Sikkim University*
*E-mail: rmukherjee@cus.ac.in*

**Abstract:** Post Modigliani and Miller (1958), the concept of usage of arbitrage created a permanent mark on the discourses of financial framework. The arbitrage process is largely based on information dissemination amongst the stakeholders operating in the financial market. The advent of the efficient market Hypothesis (Fama (1965) (1970) Makiel (1989) Fama et al. (1969) Ross (1973) Laffont & Maskin (1990) Mitchell & Mulherin (1994) Chan et al (2008) Urquhart & McGroarty (2016), draws close to the M&M hypothesis. Giving importance to the arbitrage process, which effects the price discovery in the stock market. This divided the market as random and efficient cohort system. The focus was on which information forms a key factor in deciding the price formation in the market. However, the conventional techniques of analysis do not permit the price cycles to be interpreted beyond its singular wave-like cyclical movement. The apparent cyclic measurement is not coherent as the technical analysis does not give sustained result. Hence adaption of theories and computation from mathematical methods of physics ensures that these cycles are decomposed and the effect of the broken-down cycles is interpreted to understand the overall effect of information on price formation and discovery. In order to break the cycle this paper uses spectrum analysis to decompose and understand the above-said phenomenon in determining the price behavior in National Stock Exchange of India (NSE).

**Research Methods:** This study employs computational methods to transform stock price time series data into the frequency domain using the Fourier Transform. When the data exhibit multiple frequencies, the cyclical movements become complex and necessitate a more detailed analysis. To address this, the study proposes an enhancement to the integral transformation method that will aid in identifying the factors and nature of the information present in the market.

**Outcome:** This paper presents a computational framework using an extended form of an integral transform, that can be utilized to study financial data. It analyses this phenomenon using a simulated dataset first, applicable to analyzing stock market data. Thereafter, the original tick value data set of NSE was put in comparison to the results of the simulated data. The extended integral transformation allows for a deeper and more complex examination of the stock market, enabling a more effective analysis of the cyclical data. This approach aims to improve our understanding of the underlying factors related to fluctuations and cyclical patterns in the market.

**Contribution of authors:** Professor Abhijit Dutta introduced the concept for the paper, suggesting that this approach could serve as a pioneering contribution and aid in its conceptualization and finalizing of the manuscript.

Mr. Kiran Sharma provided essential literature and theoretical background for the paper and organised the data set for the study, filtered it, and contributed to the manuscript, incorporating valuable inputs from all authors.

Dr. Rupak Mukherjee contributed to the analysis and methodology sections by employing mathematical computational approaches, offering valuable insights into technical analysis, and assisting in conceptualizing the methodology.

## 1. Introduction and Theoretical Background.

The stock market is a vital component of the economy, acting as a significant indicator of its overall health. The performance of companies, as reflected in their stock prices listed on various exchanges, is closely monitored by investors, leading to informed investment decisions. Typically, rising stock prices signal improved company performance, while declining prices suggest negative trends. Consequently, tracking stock prices is essential for assessing both individual corporate performance and broader market dynamics. Stock prices often exhibit cyclical movements influenced by changing macroeconomic factors. Investor reactions to specific economic indicators can lead to fluctuations in stock prices, Martynova & Renneboog (2008). Individual stocks display their unique cyclical patterns, which are studied either in isolation or within the context of the larger market. This variability means that while stocks may align with major cycles, they also follow their own minor cycles, Ashby (1927). It's important to recognize that cyclical stock price behaviours may arise from autocorrelation and trends rather than long-term cycles, Aydogan & Booth (1988). The phases of stock prices, shaped by market dynamics and available information, reflect how the market processes data. These cyclical phases can help predict future price movements. Analyzing the cyclical components of time series data is instrumental in understanding market trends and forecasting prices, aiding in the comprehension of both short- and long-term movements (Idrees et al., 2019). Market forecasts rely significantly on recognizing the patterns in price sequences, necessitating a thorough analysis of historical prices to enhance predictive accuracy, Cowles (1944).

Price movements are influenced by a wide array of direct and indirect factors, suggesting that fluctuations stem from changes in the market and the information at hand. This phenomenon indicates that the market efficiently assimilates all available data Fama (1970). Understanding market efficiency and the various phases of price behavior can facilitate market trend predictions, enabling investors to make informed decisions. The aspects of price fluctuations and phases also present opportunities for arbitrage—buying at lower prices and selling at higher prices with minimized risk exposure. Although achieving risk-free arbitrage is challenging, frameworks like the Capital Asset Pricing Model (CAPM) and Arbitrage Pricing Theory (APT) offer insights into how different factors influence security pricing and arbitrage opportunities, Ross (1973), Fama & French (2004). Ultimately, while numerous theories exist for market prediction, information remains a crucial asset for all market participants and plays a significant role in price discovery. Fama posited that prices are inherently random and reflect all available and private information, underscoring the importance of information in the stock market and its impact on price dynamics. Information serves as a tool for all market players—including suppliers, demand-side participants, and related stakeholders. However, disparities in information availability between firms and investors can lead to price fluctuations. Therefore, not only is the provision of information essential, but the decisions made based on this information are equally vital, necessitating regular disclosures to ensure informed market participation, Aboody & Lev (2000).

When we trace the study of price discovery and the importance of information in price discovery, we have to look back to the early 20$^{th}$ century when Bachelier (1900), in his work "The Theory of Speculation," discussed the application of probability theory to analyze prices and the profitability of forward contracts and options. Fama (1965) introduced a groundbreaking approach to understanding stock market prices, arguing that they are random and follow a random walk and laid the foundation for the Efficient Market Hypothesis, stating that stock prices reflect all available information, and since no one can access all the information in the market, it is impossible for anyone to achieve returns above average. In continuation, Fama, (1970), introduced three forms of market efficiency: Weak form efficiency suggests that all past price and volume information is fully reflected in current stock prices, Semi-strong form efficiency states that all publicly available information is reflected in prices and Strong-form efficiency asserts that both public and private information are fully incorporated into stock prices. Makiel (1989) sought to examine the validity of the EMH by conducting statistical analyses like Momentum, return reversals, seasonal effects, and predictability as well as event studies to analyze how stocks react to various events. He found that markets are not perfectly efficient; instead, they exhibit occasional predictable patterns that are not robust. Fama et al. (1969) studied 940 NYSE stock splits to assess how stock prices adjust and to test the

Efficient Market Hypothesis (EMH). They analyzed whether stock splits provide new market information and how quickly it is incorporated. Their regression and residual analyses revealed that the market anticipates splits, adjusts prices accordingly, and fully incorporates the information, thus supporting the EMH. Levy (1967) highlighted both the strengths and limitations of the random walk theory, emphasizing the need for more rigorous testing. His work paved the way for further research on market efficiency. Laffont & Maskin (1990) investigate whether the Efficient Market Hypothesis (EMH) holds under two conditions: pooling equilibrium, where prices do not reveal private information, and separating equilibrium, where prices do. They analyze how large traders can conceal information using Perfect Bayesian Equilibrium and comparative statistics. They found that while separating equilibrium exists, pooling equilibrium is preferred. This indicates that EMH may fail in imperfect competition and when large traders strategically hide information, preventing prices from fully reflecting private information. Mitchell & Mulherin (1994) examined the relationship between the information publicly reported in the Dow Jones & Company newspaper and U.S. securities market activity, which includes trading volume and market returns. By performing regression and correlation analyses, they found that news announcements are positively correlated with both trading volumes and market returns. Additionally, they discovered that news announcements have a more direct impact on trading activity than on price movements. Chan et al. (2008) investigated information asymmetry between foreign and domestic shares in the Chinese stock market. They utilized measures such as Price Impact (PI), Adverse Selection Component (AS), and Probability of Informed Trading (PIN), in addition to conducting both univariate and multivariate regression analyses. Their analyses suggested the presence of information asymmetry between foreign and domestic shares in the Chinese stock market [11]. The study by Urquhart & McGroarty (2016) found evidence of an efficient market in the S&P 500 and EUROSTOXX 50 using the variance ratio and AR-GARCH models. Their research also focused on the Adaptive Market Hypothesis, revealing that efficient markets are not always static but evolve over time.

The stock market's pricing mechanisms can sometimes reflect manipulated information, where individuals or groups may exploit unsuspecting investors to benefit from price predictions and discoveries. As more investors enter the market, the demand for information increases, leading to a rise in manipulation practices. These dynamics creates disparities among investors, contributing to market volatility Agarwal & Wu (2006). In light of the efficient market hypothesis proposed by Fama, which remains pertinent today, it's crucial to explore the origins and lifecycle of various market patterns. Makiel (1989) emphasized the need for additional research to discern whether these patterns arise from data mining or represent genuine market behavior. Effective technical analysis is vital for utilizing information wisely, offering protection to investors against adverse macroeconomic shifts. The necessity for sound regulation and honest information dissemination in the stock market is underscored by the motives and characteristics of information holders—whether they are firms, institutions, or broker groups. Ultimately, the way information is introduced into the market significantly impacts its dynamics. In discussions of price discovery, it's important to recognize that price patterns are not always symmetrical, often influenced by uncertainties and information asymmetry. The interplay of these elements has spurred interest in examining price determination and prediction, enabling more informed decision-making. Market prediction processes present opportunities for investors to strategize based on price fluctuations, particularly through arbitrage strategies that involve buying low and selling high while managing risk. Investors continually seek to capitalize on arbitrage opportunities, though avoiding all risk is unrealistic. Theories such as the Capital Asset Pricing Model (CAPM) and Arbitrage Pricing Theory (APT) provide valuable insights into the various factors that affect security prices and outline potential arbitrage prospects Ross (1973), Fama & French (2004). At the core of market fluctuations lies the availability of information; those who possess timely insights are often positioned to profit, as information shapes market behavior. Considering market efficiency and price behavior, predicting market trends becomes feasible. Analysts utilize three primary approaches to understand price variations and support informed decision-making:

## 1.2.1. Fundamental Analysis

The roots of fundamental analysis can be traced back to the 1930s, notably through the work of Williams (1930) and later popularized by Graham & Dodd (1934) in their seminal book "Securities Analysis" (1934), along with Graham's "The Intelligent Investor" (1949). These works marked the introduction of fundamental analysis to the stock market. Fundamental analysis considers a range of variables, including macroeconomic factors, industry dynamics, and company-specific elements, to determine the value of financial assets. The fundamental value calculated can be influenced by current news and information; thus, those with timely and comprehensive access to information often have an advantage over those who are less informed. However, this analysis can be complex, requiring consideration of multiple variables and significant time investment, making it more suitable for long-term investors.

## 1.2.2. Technical Analysis

In contrast, due to the desire for quicker gains and faster analysis of market trends, technical analysis emerged. This approach posits that all available information is already reflected in stock prices, allowing analysts to focus on price trends using historical data to capitalize on market movements. The foundations of technical analysis were laid out in "The Stock Market Barometer" by Hamilton (1992) and were further developed through the Dow Theory, popularized by Dow (1899) (1902). Rhea (1932) contributed to the popularity of this analytical method in his market letters and his book "The Dow Theory". To measure the theory, Dow calculates average market movements, specifically the Dow Jones Industrial Average (DJIA) introduced in 1896 and the Dow Jones Railroad Average (DJRA) also established in 1896. This implies that when the values of these two averages diverge, it serves as a warning to the market; conversely. A convergence of these, on the other hand, indicates a buying opportunity, and the opposite applies as well. This concept is referred to as technical analysis Griffioen (2003).

## 1.2.3 Computational Approach

Technical analysis has experienced significant growth, particularly among short-term traders. Due to the complexity, noisy, chaotic, volatile and vast amount of data that needs to be analyzed, machine-learning techniques have emerged and delivered superior results. The computational approach within technical analysis allows for more precise outcomes and the ability to manage extensive data sets. As a result, many analysts utilize computational and machine-learning methods to analyze financial data. The stock market data is complex and non-stationary, making traditional approaches to analysis often inaccurate, as they typically treat stock data as a linear time series. This issue is highlighted in the study by Kumar et al. (2021), which reviewed recent developments in the application of computational intelligence in the stock market. They found that methods such as Artificial Neural Networks (ANNs) and Support Vector Machines (SVMs) are increasingly being used instead of traditional techniques. The researchers conducted surveys to demonstrate the effectiveness and accuracy of computational intelligence methods in this context. Similarly, Dunis et al. (2014) provided insights into various computational techniques for analyzing the stock market. Sonkavde et al. (2023) present a comprehensive analysis of various algorithms, including time series analysis and deep learning techniques, for stock market analysis and predictions. They discuss the application of methods such as linear regression, ARIMA, and other analytical approaches within a computational framework. In the realm of technical analysis, numerous indicators exist, with candlestick and chart patterns recognized as fundamentally important and effective. Regression and image processing techniques, as highlighted by Li et al. (2020), Fauzi & Wahyudi (2016), further enhance these analyses. The findings indicate that technical analysis can yield more accurate and effective results for market prediction and forecasting. Both technical and computational approaches take into account time series data, operating under the assumption that the market is efficient, and utilize cyclical phases, patterns, and image processing techniques to evaluate the stock market. Considering the effectiveness of computational approaches compared to other methods, this study emphasizes the importance of utilizing computational analysis to explore stock market data. This study will put into the integral The Fourier Transform, particularly valued in the field of engineering and physics, is often employed to decompose time series data, primarily focusing on signals as they transition from the time domain to the frequency domain. This mathematical technique is utilized to analyze and represent signals by breaking them down into their constituent frequency components Nussbaumer (1982).

The Fourier Transform (FT) is a mathematical tool used for analyzing frequency data and decomposing complex signals into constituent frequencies, particularly useful for cyclical movements. It enables spectral analysis by providing a spectrum of a time series data, revealing dominant frequencies and patterns, such as seasonal trends or business cycles. The transformation properties allow for the superposition of signals in the frequency domain, simplifying the analysis of their interactions. The Fast Fourier Transform (FFT) is an efficient algorithm that reduces computational complexity, making it feasible to analyze large datasets. The Fourier Transform (FT) has applications across multiple disciplines, including signal processing, economics, and engineering, enhancing our comprehension of cyclical behaviours Cochran et al. (1967). This concept of Fourier transforms, along with their effectiveness in analyzing the cyclical phases of time series data, will be applied to the analysis of our NSE stock data.

Here in this study we employ FFTW library (https://www.fftw.org/) within a computational framework named TARA (https://rupakmukherjee.github.io/TARA/)(Mukherjee et al.,2018) (Mukherjee, 2019) (Mukherjee & Ganesh, 2018). The TARA simulation architecture is a multi-dimensional pseudo-spectral solver to simulate weakly compressible and incompressible fluid or plasma turbulence (Mukherjee et al., 2019) (Mukherjee et al., (2019) (Saikia & Mukherjee 2024). TARA is flexible for adding higher order fluid-moment equations with minimal computational overhead (Mukherjee et al.,2019) (Mukherjee et al., 2018). This framework runs efficiently on CPU as well as GPU architecture (Mukherjee & Ganesh, 2019) (Mukherjee et al., 2019). In addition, the performance scales efficiently under MPI on massively parallel shared- and distributed-memory computers (Biswas et al.,2021) (Biswas et al., 2020) (Gupta et al.,2019). Here in this work, we extend the TARA diagnostic module and perform bispectrum analysis (Kim & Powers, 1979) of time-series data. We benchmark our new module with existing analytical results and then perform analysis of NSE data with our new modules.

## 2. Research gap

In reviewing the literature on stock data analysis, it became evident that the application of computational approaches, particularly those derived from mathematical physics, is surprisingly limited. This is noteworthy, given that such methods often yield superior results when examining price movements. The effectiveness of decomposing time series data into the frequency domain can enhance our understanding of the relationship between different frequencies, thereby providing a deeper insight into market asymmetry and the various factors influencing the market. However, this approach has not been extensively explored in existing studies.

## 3. Research objective

Considering the aspects of information asymmetry and market efficiency, this study aims to contribute to advancements in financial market analysis, particularly regarding how asymmetric information affects market behavior. The primary objective of this paper is to explore the implications of information and the underlying causes of price movements and their phases in the stock market. This study will provide a comprehensive computational framework that integrates concepts and techniques of mathematical physics especially the integral transforms, this study aims to provide a distinct and effective approach to examining stock market dynamics and their influencing factors. Ultimately, seeks to enhance our understanding of the phases and interactions between different cyclical modes and contribute to the discourse on information dynamics within the financial sector.

## 4. Proposition:

1. Stock prices show a cyclical pattern of movement. The movements are repetitive and hence a pattern analysis can lead to prediction of the market price rallies. Peters (1996) Sharma & Wongbangpo (2002) Akar & Bakaya (2011) Cootner (1962) (1986) Hirsh (2012)

2. If the price rallies are independent then some cyclical patterns will be formed in sequence. However, if a bias/information is planted through some artificial pattern, then a superposition of such frequency will be generated, with a surrogated 'phase information' hidden inside it.
3. Since standard integral transformations (for example Fourier-transformation) cannot distinguish the patterns generated from genuinely independent frequencies, and the patterns generated from the independent as well as biased frequencies, there is a chance that the biased frequencies have been planted artificially. This will make the market predictive for the person who plants the bias artificially.
4. In this work, we propose an extended form of an integral transformation to identify the surrogated 'phase information' due to artificially planted information from the raw stock data, if there is any.

## 5. Research Methods

This study reports a new data analysis method to analyze phase-relationships between different frequencies. The importance of such method is described in the Proposition section. We provide a short tutorial of our analysis tool with some illustrative data first. Then we apply our tool for real stock market data.

Given a function in time domain, it is possible (given that it satisfies Dirichlet criteria) to convert it to frequency domain by taking a Fourier transform of the given function.

$$F(\omega) = \frac{1}{\sqrt{2\pi}} \int_{-\infty}^{\infty} f(t) e^{i\omega t} \, dt \qquad \ldots\ldots\ldots\ldots\text{Equation 1}$$

Here f(t) denotes the function in time domain and the function F(ω) denotes the corresponding Fourier transform in frequency (ω) domain. We can extend this integral transform into cases of discrete datasets as well in the following way:

$$F(\omega) = \sum_{t=0}^{N-1} f(t) e^{-i 2\pi \omega t / N} \qquad \ldots\ldots\ldots\ldots\text{Equation 2}$$

The amplitudes of the Fourier modes denote the strength of the corresponding frequency present in the timeseries data (f(t)). Each one of the frequencies represent one cyclical pattern movement of the data. When there are multiple frequencies present, the cyclical movement becomes cumbersome and requires an in-depth analysis. In our analysis we use a numerical library: FFTW (Fastest Fourier Transform in the West) to evaluate the Fourier transform of both illustrative data as well as the stock market data.

However, the discrete Fourier transform of data (in real numbers) generate Fourier modes in complex number domain. It is usual practice to evaluate the square norm of the complex data to draw inferences, as a square norm maps the data back to the real-number space. However, this method throws away the phase relations between different complex Fourier modes and hence we may potentially loose the valuable information about the phases and frequencies between the modes which are hidden inside the data.

In this analysis, we propose an extension of this integral transform (in this report, Fourier transform) such that the phase relations are retained. Thus, an artificially planted information/frequency will appear as a hotspot in our extended integral transform, as it will maintain a fixed phase relationship with the naturally occurring cluster of independent frequencies.

### 5.1. Tools and Techniques:

To begin with, we'll create an illustrative sinusoidal dataset that has a fixed frequency and a fixed phase. This study will then introduce another sinusoidal dataset with a different fixed frequency and phase. The sinusoidal datasets are generated using a sine wave, which is a repeating wave pattern often described as oscillation. We create two copies of this same dataset. In one of the copies, further add another sinusoidal dataset of a superposition of the two frequencies and a third fixed phase. This third phase is independent of the earlier two phases.

$$f(t) = \cos(\omega_\alpha t + \theta_\alpha) + \cos(\omega_\beta t + \theta_\beta) + \cos(\omega_\gamma + \theta_\gamma), \quad \frac{1}{\omega_\gamma} = \frac{1}{\omega_\alpha} + \frac{1}{\omega_\beta}, \quad \theta_\gamma, \theta_\alpha, \theta_\beta \text{ are all random phases.}$$

In the other copy of the dataset, the study will add a sinusoidal dataset of a superposition of the two frequencies and a third phase, which is the addition of the earlier two phases.

$$f(t) = \cos(\omega_\alpha t + \theta_\alpha) + \cos(\omega_\beta t + \theta_\beta) + \cos(\omega_\gamma + \theta_\gamma), \quad \frac{1}{\omega_\gamma} = \frac{1}{\omega_\alpha} + \frac{1}{\omega_\beta}, \quad \theta_\gamma = \theta_\alpha + \theta_\beta, \quad \theta_\alpha, \theta_\beta \text{ are random phases.}$$

Now, the study will plant some low-amplitude artificial noise in both datasets. However, this addition of noise is optional and it only indicates the robustness of the diagnostic technique we follow. Thus, both the illustrative sample data consist of an amalgamation of three different cycles, which is hard to distinguish by a quick specular inspection. The first dataset does not possess any fixed phase relation between the different cycles, while the second one, by construction, contains a phase relationship between the three modes. Now let us take the Fourier transform of these two datasets. The Fourier transform will provide amplitudes in the complex number domain. To draw any meaningful conclusion, we will have to calculate the modulus value (the square norm) of the amplitudes. This will indicate the strength of each of the cycles in the data. However, this also throws away the phase relations that were otherwise present in the complex amplitudes. Thus, a regular Fourier transform of the two datasets will look identical and the information that the superposed frequency had a pre-specified phase relationship in the second dataset can never be retrieved from this analysis.

Here in this study, we propose an extended form of integral transform (to be specific, an extended Fourier transform) that is capable of retaining this phase relationship. This method has been explored earlier in the context of electrical engineering and some turbulence-related studies in astrophysics as well as plasmas in nuclear fusion reactors. The method is simple to implement and easy to evaluate. First, the study will calculate the complex conjugate of the Fourier modes. Then multiply two different complex Fourier modes and further multiply them with the complex conjugated Fourier mode where the mode number is the addition of the previous two modes. This study will further sum over all such possible combinations. This new quantity retains the phase information of all the modes. This can be easily tested from the two illustrative datasets that was prepared earlier. The first dataset will not show any correlation, while the second one will now show spikes for those modes which has been used to create the third mode. *(Refer Annexure 1)*

### 5.2. Extended Fourier Transform

In this study, we will construct a new quantity as

$$P(\omega_\alpha, \omega_\beta) = F(\omega_\alpha)F(\omega_\beta)F*(\omega_\alpha + \omega_\beta) \quad \ldots\ldots\ldots\ldots\text{Equation 3}$$

Here we multiply two different Fourier component $F(\omega_\alpha)$ and $F(\omega_\beta)$ and then multiply the result with the complex conjugate of Fourier modes at frequency $(\omega_\alpha + \omega_\beta)$

NOTE: By undergoing the complex conjugate of the Fourier modes when FT is carried out a function (wave or signal) we express it as a sum of sinusoidal components (wave) of different frequencies these components are Fourier modes) we are changing the direction of the wave to retain the phase information.

Multiplying the two complex Fourier modes $F(\omega_\alpha)$ and $F(\omega_\beta)$ represents the interaction between the frequency component $\omega_\alpha$ *and* $\omega_\beta$. When we multiply the two complex modes with the complex conjugate mode, $F*(\omega_\alpha + \omega_\beta)$, this will ensure that the new quantity $P(\omega_\alpha, \omega_\beta)$ retains the phase information or connections of the interacting modes. If strong phase interaction exists between $\omega_\alpha$ and $\omega_\beta$ it will appear as a strong correlation in $P(\omega_\alpha, \omega_\beta)$.

Let us consider this extended Fourier transform in the earlier two data sets that we have prepared. The 1st data set (shows no correlation), which means if there is no strong interaction between different modes then $P(\omega_\alpha, \omega_\beta)$ will not show significant values which means that mode interactions are weak or do not exist. 2nd data set (shows correlated modes), which means that if certain modes interact to form the third mode, i.e. $F(\omega_\alpha)$ and $F(\omega_\beta)$ contribute to $F(\omega_\alpha, \omega_\beta)$, then $P(\omega_\alpha, \omega_\beta)$ will show peaks for those interactions which is because the phase relations are preserved. *(Refer Annexure 1)*

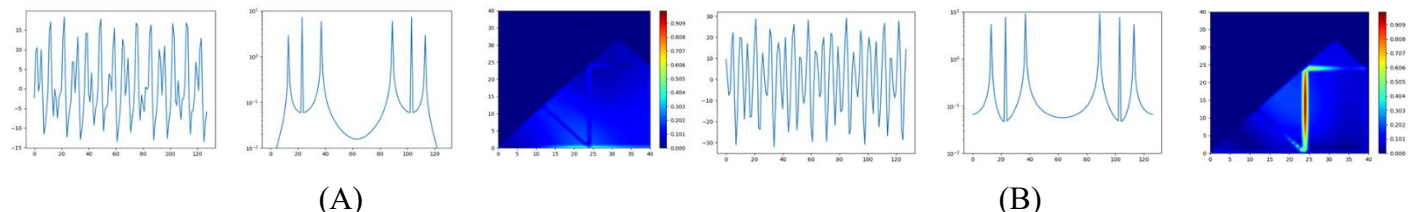

(A)          (B)

*Figure 1: FFT (Fastest Fourier Transform) and BSP(Bispectrum) of the First and Second data sets. A is for the First Data set and B is for the Second data set.*

### 5.3. Period of study and nature of data

This study uses secondary data from the stocks listed in NSE, the study uses tick data o 1-minute interval from 2015 to 2022. The data is collected from 2 companies, each in 5 different industries, on the National Stock Exchange as provided below.

| BANKING | FMCG | AUTOMOBILE | IT | IRON & STEEL |
|---|---|---|---|---|
| ICICI | Dabur | Mahindra & Mahindra | Infosys | JSW Steel |
| SBI | HUL | Tata Motors | Tata Consultancy | Tata Steel |
| NIFTY 50 INDEX DATA (2015-2022) | | | | |

This study uses unaltered or raw data from the above-mentioned companies under different industries so that a precise and accurate study can be conducted.

## 6. Results and Discussion

Utilizing the extended Fourier transform to analyze two different simulated data sets reveals distinct outcomes when visualized. In the first data set, where all Fourier modes are independent, no spikes are observed, indicating a lack of correlation between the diverse modes. Conversely, when the third mode is generated using the frequency components of the first two modes, spikes are observed, signalling a relationship between the modes. This method of transitioning data from the time domain to the frequency domain and examining various frequency modes is essential for understanding the interdependence among different Fourier modes in stock market time series data. When applying these extended Fourier methods to the stock market data, we found the following results:

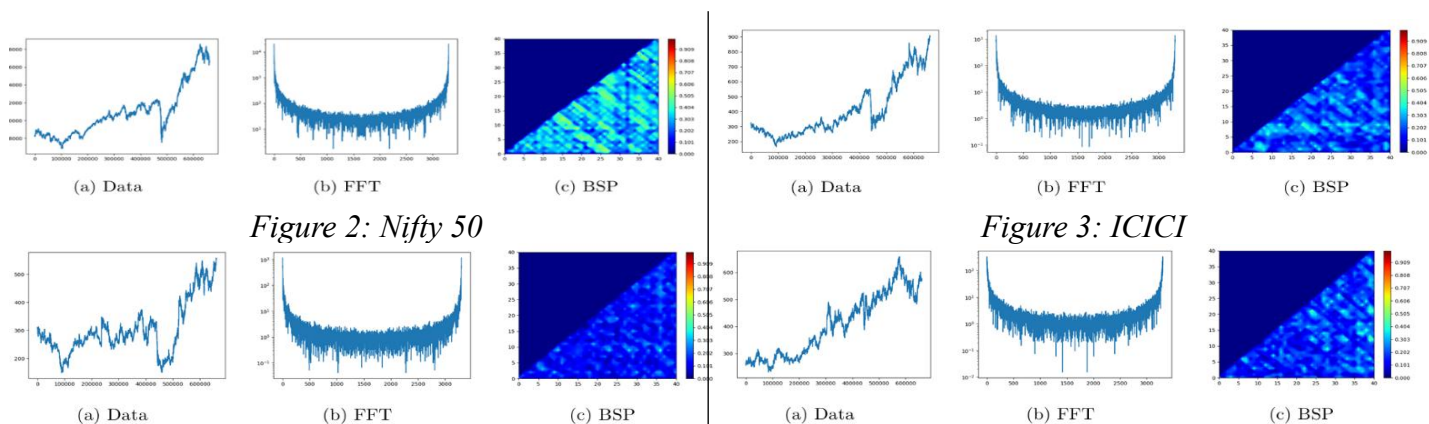

*Figure 2: Nifty 50*          *Figure 3: ICICI*

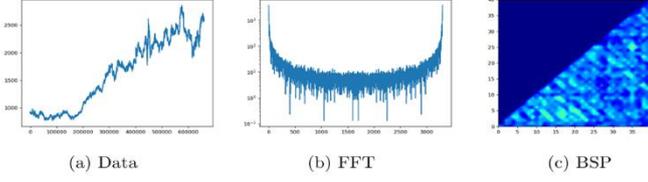

Figure 4: SBI

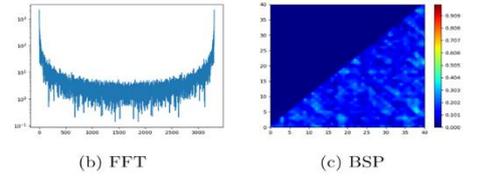

Figure 5: Dabur

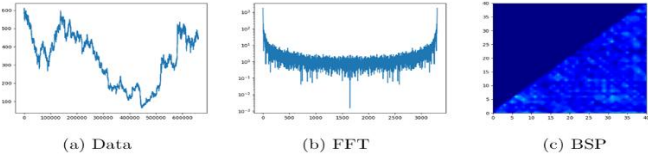

Figure 6: HUL

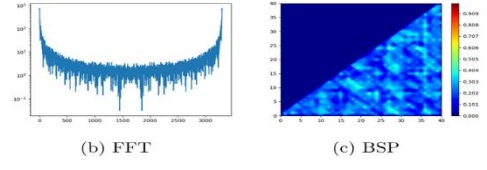

Figure 7: MM

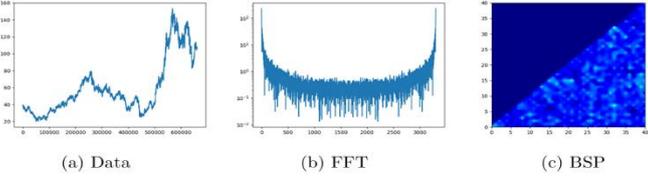

Figure 8: Tata Motors

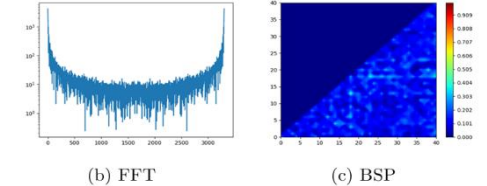

Figure 9: JSW Steel

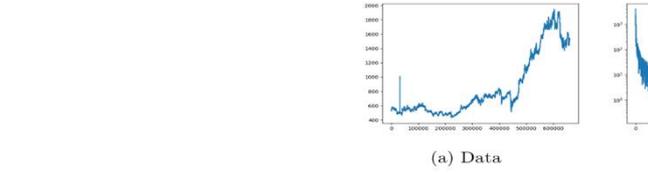

Figure 10: Tata Steel

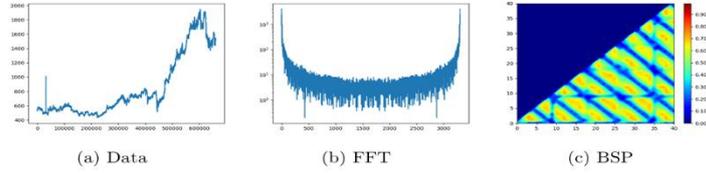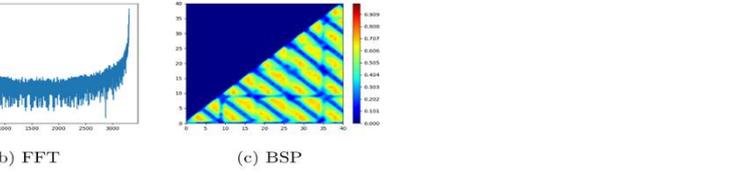

Figure 11: Tata Consultancy

Figure 12: Infosys

Analyzing stock data from the above-mentioned perspectives yields a wealth of information, allowing us to draw numerous conclusions and assumptions. The second dataset provides insights into phase relations, exhibiting spikes as shown in Figure 1(B). A similar trend can be observed in the stock market data for Infosys, illustrated in Figure 12. These spikes indicate interactions and relationships between the various frequency modes when analyzed using the Extended Fourier Transform (EFT). However, in other stock datasets, we do not observe such interactions between the two modes to create a third mode. Our analysis of stock market data using the Extended Fourier Transform focused on retaining the phase relations among different Fourier modes. This method reveals spikes when visualized. In the case of Infosys, we identified spikes that signify an interplay within the Fourier modes of its data cycle. In contrast, other stock prices do not exhibit such relationships, as seen in the first dataset. Furthermore, the Infosys dataset shows results consistent with the second dataset, where the phases and modes are interdependent. This implies that the cyclical pattern of Infosys is not independent. It suggests that factors affecting the stock market may have influenced the price fluctuations in Infosys. This raises the possibility that information may have been artificially introduced into the price of Infosys stock either publicly or privately leading to this observed behavior.

## 7. Conclusion

This article analyses the stock market and major stock prices using a mathematical physics approach, specifically the Fourier transform, which is commonly used in the field of physics. This study proposes a mathematical model called the Extended Fourier Transform (EFT). Firstly, we tested this model on two different sample datasets and then extended its application to stock market data, which included ten stock prices and the Nifty 50 index. At first when applied to the sample data set, we found the correlation between the modes on the second sample data set but no such inference can be drawn from the other sample data set. The same extended Fourier Transform which is capable of retaining the phase relations has been applied to the stock market. Our findings revealed some correlation between the phases of Infosys stock. When analyzed the frequency domain of the Infosys prices it was found that the phases which was decomposed from the Infosys data revelled some correlation between them implying that the prices of Infosys have been influenced by some environment as well as market factors. however, the results for the other stock market data were different and showed no such correlation. This suggests that there may be specific information influencing the behavior of Infosys. Furthermore, this study offers an alternative approach to analyzing financial data using concepts from mathematical physics.

**Annexure I**

1. FOURIER TRANSFORM((FT)

$$F(\omega) = \frac{1}{\sqrt{2\pi}} \int_{-\infty}^{\infty} f(t)e^{i\omega t} dt \quad \ldots\ldots\ldots\ldots\ldots \text{Equation 1}$$

Here f(t)= original function / signal.

F(ω)= Result of Fourier transform which tells us how much of each frequency (ω) is present in the signal

$e^{i\omega t}$ is a way of representing waves (it combines sine & cosine waves using complex numbers (a+ib, where a and b are real number and i is imaginary unit)

This equation helps us to see the "ingredients" Frequencies (how fast the wave comes. e.g. 1Hz =1 wave per second/ how many waves pass a point in a second) that make the signal. example if a signal consists of a mix of fast and slow wave then the FT will show both the fast and slow.

1.1. DISCRETE FOURIER TRANSFORM (DFT)

$$F(\omega) = \sum_{t=0}^{N-1} f(t)e^{-i2\pi\omega t/N} \quad \ldots\ldots\ldots\ldots\ldots \text{Equation 2}$$

F(t) is our data at real-time (stock price), F(ω) tells us how much of each frequency(ω) is present in the data.

The amplitudes (height of the wave) of the Fourier modes (FT- express function as a sum of sinusoids with different frequencies, amplitudes, and phases)- these individual sinusoids are the Fourier modes. signifies the strength of the corresponding frequency which is present in time series data. And each one frequency represents the cyclical pattern movement of data. When there is multiple frequencies present, the cyclical movement becomes unmanageable, so it requires an in-depth analysis. Hence, we use FFTW to evaluate the FT of both the illustrative data and stock market data.

When we use DFT on stock data, it generates Fourier modes in the complex number domain. We use square norm/ modules of those complex data to draw conclusions, but the square norm, which maps the data to the real number, throws away the phase relation (*where the wave starts*), which as a result may lead to losing the important information hidden inside the data.

This study focusses on providing an extension to this transformation (Fourier Transformation) such that the phase relations are retained

1.3. DETAILED TOOLS & TECHNIQUES

First, we create a sinusoidal data set of fixed frequency and fixed phase $f(t) = \cos(\omega_a t + \theta_\alpha)$ we add another sinusoidal data set of fixed but different frequency and fixed phase than the first one. $\cos(\omega_\beta t + \theta_b)$ which will be $f(t) = \cos(\omega_\alpha t + \theta_\alpha) + \cos(\omega_\beta t + \theta_\beta)$ And we will create two copies of the same data set.

❖ 1st data set: which has no phase relations, i.e. $\theta_\gamma, \theta_\alpha, \theta_\beta$ are all random and independent phases.

We will add a third sinusoidal wave with frequency $\omega_\gamma$ and an independent random phase $\theta_\gamma$. And the first data set is given below

$$f(t) = \cos(\omega_\alpha t + \theta_\alpha) + \cos(\omega_\beta t + \theta_\beta) + \cos(\omega_\gamma + \theta_\gamma),$$

Where $\omega_\gamma$ follows the relation $\frac{1}{\omega_\gamma} = \frac{1}{\omega_\alpha} + \frac{1}{\omega_\beta}$ (NOTE: phase $\theta_\gamma$ is independent of $\theta_\alpha$ and $\theta_\beta$ (phases)

❖ **2nd Data set: Fixed phase relation, i.e. $\theta_\gamma = \theta_\alpha + \theta_\beta$**

Instead of choosing an independent phase $\theta_\gamma$ We propose a fixed-phase relation that is $\theta_\gamma = \theta_\alpha + \theta_\beta$ And the second data set is given below

$$f(t) = \cos(\omega_\alpha t + \theta_\alpha) + \cos(\omega_\beta t + \theta_\beta) + \cos(\omega_\gamma + \theta_\gamma)$$

Where $\frac{1}{\omega_\gamma} = \frac{1}{\omega_\alpha} + \frac{1}{\omega_\beta}$, $\theta_\gamma = \theta_\alpha + \theta_\beta$ (NOTE phase $\theta_\gamma$ is not independent)

The above two data sets consist of a joint nature of three different cycles and are very hard to easily distinguish these three cycles as the first data set does not have any fixed phase relation, but the second one has fixed phase relations.

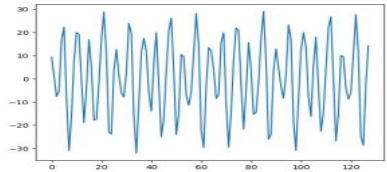
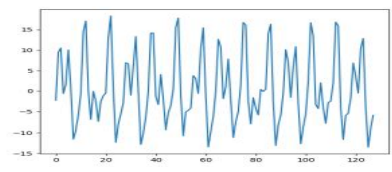

First Data set　　　　　Second Data set

We will further take the Fourier transform of the data sets. The Fourier transformation will provide amplitudes (height of the wave/ maximum value of the wave) in the complex number domain.

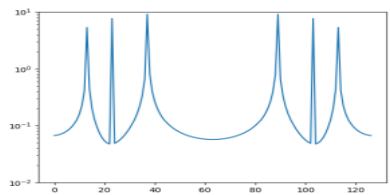
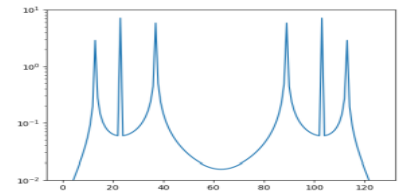

FFT of First data Set　　　　　FFT of Second data set

Conducting the Fourier transform of the data set it is unable to distinguish the data set as both shows similar symmetric information and is unable to draw any conclusions se we try to draw modulus/ square norm {which tells us the strength of each cycle of the data or power of each frequency or strength of waves or signals (Euclidean norm)} of the amplitudes. But performing square norm discards the phase information that was present in the complex amplitudes, as a result, both the data sets will look identical in the regular FT even if the information of a pre-specified phase relationship cannot be ascertained. Due to this reason, we propose an extended Fourier transform.

2. EXTENDED FOURIER TRANSFORM

In this study, we will construct a new quantity as

$$P(\omega_\alpha, \omega_\beta) = F(\omega_\alpha)F(\omega_\beta)F*(\omega_\alpha + \omega_\beta)$$

Here we multiply two different Fourier components $F(\omega_\alpha)$ and $F(\omega_\beta)$ and then multiply the result by the complex conjugate of the Fourier modes at the frequency $(\omega_\alpha + \omega_\beta)$. NOTE: By undergoing the complex conjugate of the Fourier modes (when FT is carried out on a function (wave or signal), we express it as a sum of sinusoidal components (wave) of different frequencies, these components are Fourier modes) we are changing the direction of the wave to retain the phase information. Multiplying the two Fourier modes $F(\omega_\alpha)$ and $F(\omega_\beta)$ represents the interaction between the frequency component $\omega_\alpha$ $and$ $\omega_\beta$

When we multiply the two modes with the complex conjugate $F*(\omega_\alpha + \omega_\beta)$, This will ensure that the new quantity $P(\omega_\alpha, \omega_\beta)$ retains the phase information or connections of the interacting modes. If a strong interaction exists between $\omega_\alpha$ $and$ $\omega_\beta$ it will appear as a strong correlation in $P(\omega_\alpha, \omega_\beta)$

Let us consider this extended Fourier transform in the earlier two data sets that we have prepared

The 1st data set (shows no correlation), which means if there is no strong interaction between different modes, then $P(\omega_\alpha, \omega_\beta)$ will not show significant values, which means that mode interactions are weak or do not exist.

On the 2nd data set (shows correlated modes), which means that if a certain mode interacts to form the third mode, i.e. $F(\omega_\alpha)$ and $F(\omega_\beta)$ contribute to $F(\omega_\alpha, \omega_\beta)$, then $P(\omega_\alpha, \omega_\beta)$ will show peaks for those interactions. Which is because the phase relations are preserved.

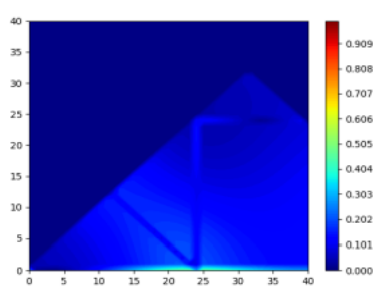 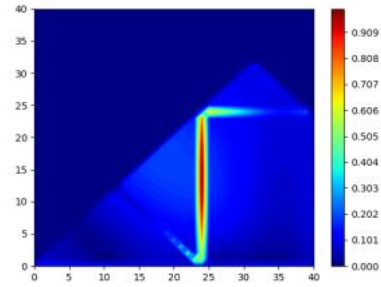

BSP of First data set             BSP of Second data set

Turbulence can be viewed as a combination of various independent cycles / frequencies present in a system. Typically, in fluid turbulence, fluid medium can sustain several independent frequencies at the same time together. This happens as fluids cannot sustain strong shear forces and fluid elements try to thermalise through conduction as well as convection mechanism. When energy is pumped into a fluid medium, typically the energy resides in large wavenumbers or in long-time cycles of fluid's motion. However, this energy cascades into shorter wavenumbers or in short-time cycles soon. This cascading of energy into shorter scales are called inertial-range of turbulence. Finally, the energy is drained out in the form of heat, through tiny viscous scales of the system. Thus, typically a fully-developed turbulent fluid contains three clear scales – the first being the scale at which the energy is being injected into the system, the second is the inertial scale where the energy cascades to the shorter scales and the third is the viscous scale where the energy gets drained out from the medium in the form of heat/radiation. The second phase, that is, the inertial scale plays the key role in developing the turbulence in a fluid. This inertial scale transfers the energy (/amplitude in stock market) from large scale vortices (/long-time cycles) to smaller vortices (/short-time cycles or fluctuations) through some nonlinear interactions. It has been found that in a fully developed turbulence, these nonlinear interactions between different modes (/cycles) actually retain information from both amplitude as well as the phase of each of the modes (/cycles). Thus, merely an analysis using the energy spectrum falls short to understand the actual nature of the turbulence present in the system. For example, inside a fully developed turbulence, we expect there will be no correlation between the individual phases and thus each cycle will have its own unique phase information. In addition, if there is a huge correlation between the phases of individual cycles, one can conclude that, the medium has not reached into a fully developed turbulent phase. In this report we describe a method to dig out this phase information from a stock market data (using an integral transform method) and check for the quality or degree of turbulence in these individual data. This method is already in use to identify and quantify the level of turbulence in fluid as well as in plasmas (Kumar, 2019) (Kumar, 2018). However, to the best of our knowledge this is the first time we employ this technique to categorise turbulence in stock market data.